\documentclass[aps,prd,floatfix,superscriptaddress,twocolumn,preprintnumbers]{revtex4} 
\usepackage{amssymb}
\usepackage{amsmath}
\usepackage{amsfonts}
\usepackage{epsfig}             
\usepackage{graphicx}
\usepackage{stackrel}
\usepackage{tabularx}
\usepackage{color}
\usepackage{dsfont}
\usepackage[T1]{fontenc}
\usepackage{natbib}

\usepackage{lipsum}

\definecolor{darkerblue}{rgb}{0.0,0.0,0.5}
\newcommand{\seq}{\begin{subequations}}
\newcommand{\sen}{\end{subequations}}

\newcommand{\eq}{\begin{eqnarray}}
\newcommand{\en}{\end{eqnarray}}

\def\nn{\nonumber}

\begin{document}
	
	\title{ Lepton flavor violating  dark photon} 

	\author{Alexey~S.~Zhevlakov \footnote{{\bf e-mail}: zhevlakov@theor.jinr.ru}} 
\affiliation{Bogoliubov Laboratory of Theoretical Physics, JINR, 141980 Dubna, Russia} 
\affiliation{Matrosov Institute for System Dynamics and 
	Control Theory SB RAS, \\  Lermontov str., 134, 664033, Irkutsk, Russia } 

\author{Dmitry V.~Kirpichnikov\footnote{{\bf e-mail}: kirpich@ms2.inr.ac.ru}}
\affiliation{Institute for Nuclear Research of the Russian Academy 
	of Sciences, 117312 Moscow, Russia} 
		
	\author{Valery~E.~Lyubovitskij \footnote{{\bf e-mail}: valeri.lyubovitskij@uni-tuebingen.de }} 
	\affiliation{Institut f\"ur Theoretische Physik, Universit\"at T\"ubingen, \\
		Kepler Center for Astro and Particle Physics, \\ 
		Auf der Morgenstelle 14, D-72076 T\"ubingen, Germany} 
	\affiliation{Departamento de F\'\i sica y Centro Cient\'\i fico
		Tecnol\'ogico de Valpara\'\i so-CCTVal, \\ 
		Universidad T\'ecnica Federico Santa Mar\'\i a, Casilla 110-V, Valpara\'\i so, Chile}
	\affiliation{Millennium Institute for Subatomic Physics at
		the High-Energy Frontier (SAPHIR) of ANID, \\
		Fern\'andez Concha 700, Santiago, Chile}
	
	\date{\today}
	
	\begin{abstract}
We study the possible impact of dark photons on lepton flavor phenomenology.
We derive the constraints on non-diagonal dark photon couplings with leptons by analyzing corresponding  contributions to lepton anomalous magnetic moments, rare lepton decays and the prospects of fixed-target experiments aiming for search
for light dark matter based on missing energy/momentum techniques.  
	\end{abstract}
		
	\maketitle
	
	\section{Introduction}
 \label{Intro}
	
 Direct search for dark matter (DM) remains one of the most  
 challenging issues in particle physics.  
 Astrophysical data and cosmological observations at different scales
 imply the indirect evidence of DM. Despite numerous intensive direct searches for DM in the accelerator based experiments, little is known about origin and dynamics of weakly coupled
 particles of the hidden sector. In addition, the muon $(g-2)$ anomaly~\cite{Aoyama:2020ynm} and recent tensions between Standard Model (SM) expectation and experimental
measurements~\cite{XENON:2020rca,CDF:2022hxs,MiniBooNE:2020pnu,%
Krasznahorkay:2015iga}  have been stimulated a development of various beyond SM (BSM) scenarios involving sub-GeV hidden sector
particles~\cite{Antel:2023hkf}.

Typically such scenarios imply a feebly interacting mediator (portal) states connecting the BSM sector with SM particles.
In particular, recently several such hidden sector scenarios  
have been discussed in literature: the Higgs portal~\cite{Arcadi:2019lka,Gninenko:2022ttd,Davoudiasl:2021mjy},
the tensor
portal~\cite{Voronchikhin:2023znz,Voronchikhin:2022rwc,Kang:2020huh},
the dark photon portal~\cite{Fortuna:2020wwx,Buras:2021btx,Kachanovich:2021eqa}, sterile neutrino
portal~\cite{Escudero:2016tzx},
axion portal~\cite{Nomura:2008ru},
and Stueckelberg
portal~\cite{Kachanovich:2021eqa,Lyubovitskij:2022hna}. 

It is worth noticing that some hidden sector models suggest
an idea of lepton non-universality and lepton flavor violation (LFV).  In this sense, a light sub-GeV hidden particles may potentially explain a muon $(g-2)$ anomaly and other SM tensions in particle physics implying  LFV effects~\cite{Arcadi:2017xbo,Han:2020dwo,Aloni:2017ixa,Poh:2017tfo,Araki:2022xqp}. We note that  neutrino oscillations provide clear experimental evidence for LFV, however for the charged lepton sector these effects are strongly suppressed. Therefore, in order to probe LFV phenomena one may address a new light vector field that violates charged lepton flavor at tree level. To be more specific, in the present paper, we discuss the examination of dark photon portal that can be relevant
for LFV lepton decays and LFV processes at fixed target experiments. 

In case of dark photon, which can acquire a mass via the Stueckelberg mechanism, the Lagrangian of its interaction with DM can be written as follows~\cite{Kachanovich:2021eqa}
	\begin{eqnarray}\label{eq:DS-Lagr-gf}
	{\cal L}'_{\rm DS} &=&  - 
	\frac{1}{4} \, {A}_{\mu\nu}' {A}^{\prime \mu\nu} 
	\,+\, \frac{m_{A'}^{2}}{2} A'_{\mu} A^{\prime \mu} \nonumber\\
	\,&+&\, \bar\chi \, (i\not\!\!D_{\chi} - m_\chi) \, \chi   
	-  \frac{1}{2\xi}\left(\partial_{\mu}A^{\prime \mu} \right)^{2} \nonumber\\
	&+& \frac{1}{2} \partial_{\mu}\sigma\partial^{\mu}\sigma - \xi \frac{m_{A'}^{2}}{2} \sigma^{2} \,,
	\end{eqnarray} 
where $m_{A'}$ is the mass of dark photon, $\chi$ is 
a Dirac dark matter field, $\sigma$ is the singlet Stueckelberg field, $\xi$ is the gauge-fixing parameter.
The interaction of dark photon, $A_\mu'$, with charged leptons, $\psi_i$,  can include both diagonal and non-diagonal couplings	
	\begin{eqnarray}\label{eq:A'-psi-1} 
	{\cal L}_{\rm A'\psi}&=&  \sum_{i,k=e,\mu,\tau} A'_{\mu}  \bar{\psi}_i \gamma^{\mu}\left(g^{V}_{ik} 
	+ g^{A}_{ik}\gamma_5 \right) \psi_k \,,
	\end{eqnarray}
 where $g_{ik}^V$ and $g_{ik}^A$ are the vector and axial-vector dimensionless couplings, respectively. 
Such couplings naturally arise in the 
familon scenarios~\cite{Buras:2021btx,Kachanovich:2021eqa} 
that imply an ultra-violet completion of the models. 
The bounds on  leptophilic non-diagonal, $i \neq k$, dark photon  couplings  in Eq.~(\ref{eq:A'-psi-1}) are derived 
explicitly in Refs.~\cite{Buras:2021btx, Kachanovich:2021eqa} from  
anomalous $(g-2)$ magnetic moments of charged leptons and the experimental  constraints on rare $l_i \to \gamma \, l_k $ and 
$l_i \to 3\, l_k$  decays. It is worth noticing, that 
diagonal couplings, $i=k$, in Eq.~(\ref{eq:A'-psi-1}) may induce at the one loop level the kinetic mixing between SM photon and dark photon~\cite{Holdom:1985ag} (for a recent review on the 
corresponding constraints see, e. g., Ref.~\cite{Antel:2023hkf} 
and references therein). 

The paper is organized as follows. In Sec.~\ref{AMM_lepton} we  discuss the contribution of dark photon with 
non-diagonal lepton vertices to the anomalous magnetic moment of leptons. In Sec.~\ref{LFV-two-body-decay} 
we briefly discuss two body LFV decays involving a sub-GeV vector 
in the final state, $l_i \to l_f + A'$. In 
Sec.~\ref{FixTarget}, we derive the limits for non-diagonal coupling of dark photon from fixed target 
experiments. Finally, in Sec.~\ref{summary} 
we present our bounds to dark photon non-diagonal couplings with charged leptons and discuss a prospects to search for LFV 
conversions $e N \to\mu N A'$, $\mu N \to e N A'$, 
$e N \to \tau N A'$, and $\mu N \to \tau N A'$
at the fixed target facilities.
\section{Lepton $(g-2)$ tensions} 
\label{AMM_lepton}

\begin{figure}[b]
	\includegraphics[width=0.35\textwidth, trim={2cm 22.5cm 10.5cm 1.5cm},clip]{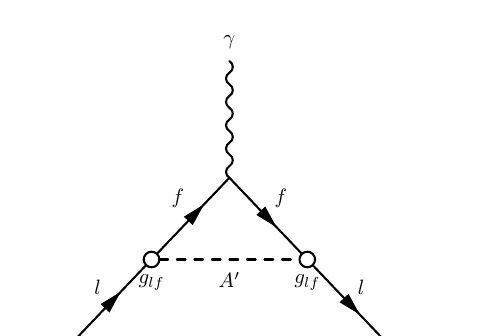}
	\caption{Diagrams describing the contributions of  
		the dark photon $A'$ to the anomalous magnetic moments 
		$\delta a_{l}$ of the leptons with taking into account flavor non-diagonal ($l\neq f$)couplings, 
		where  $l,f=e,\mu, \tau$.}
	\label{DiagG2}
\end{figure}
 
Anomalous magnetic moments of both muon or electron are 
the quantities  that can be used to constrain the 
parameters of the New physics model. In particular, 
the current discrepancies of $(g-2)$ quantities for electron and muon between the experimental measurement and theoretical prediction in the framework of SM are
\eq
\Delta a_\mu &=& (2.51 \pm 0.59) \times 10^{-9}  \qquad \text{\cite{Aoyama:2020ynm}}, 
\\
|\Delta a_e | &=& (4.8 \pm 3.0) \times 10^{-13} \qquad \text{\cite{Morel:2020dww}}. 
\en
Here we use the difference between theory and experiment for $(g-2)$ of muon based on theoretical analysis from \cite{Aoyama:2020ynm}. Now, situation with discrepancy of $(g-2)$ muon is dramatically due to the emergence of new data of two pion contribution to hadronic vacuum polarization (HVP) from CMD-3 Collaboration \cite{CMD-3:2023alj}, new measurements of $(g-2)$ from Muon g-2 experiment at Fermilab \cite{Muong-2:2023cdq} and huge data of new theory discussions about value for HVP term \cite{RBC:2023pvn,Davier:2023fpl,Borsanyi:2020mff}. 
In the BSM framework, the deviation of the magnitudes of muon and electron $(g-2)$ can be potentially explained due to the sub-GeV boson feebly interacting with leptons~\cite{Kirpichnikov:2020tcf,NA64:2021xzo}, that implies 
keeping non-zero diagonal couplings, $g^{A,V}_{ii} \neq 0$, and vanishing of the non-diagonal terms, $g^{A,V}_{ik} \equiv 0$. 
The precision of the measurements of the tau lepton anomalous magnetic moment is significantly worse, 
due the lack of experimental data on the short lived tau~\cite{Eidelman:2007sb,Workman:2022ynf}. 
For completeness, we cite Ref.~\cite{Aoyama:2020ynm} for current status of the $(g-2)$ muon puzzle not implying the BSM interpretation. 

\begin{figure}[h!]
		\includegraphics[width=0.47\textwidth, trim={0cm 0cm 0cm 0cm},clip]{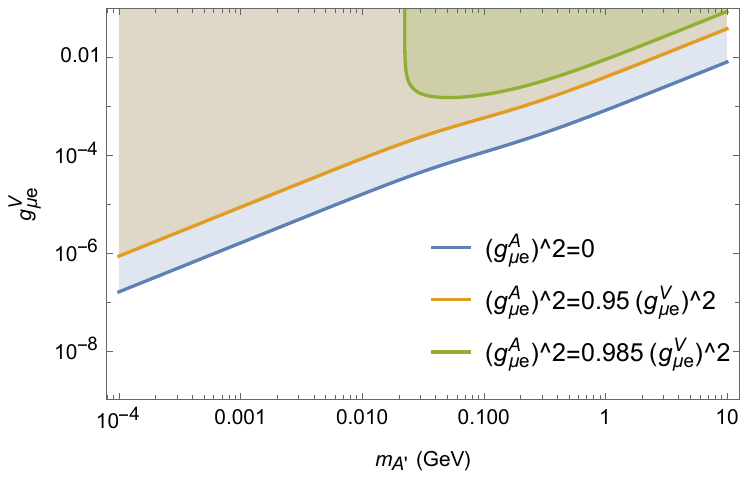}
		\caption{ Bounds for a non-diagonal coupling $g^V_{\mu e}$ from $(g-2)_e$ at different ratio between axial-vector $g^A_{\mu e}$ and vector coupling $g^V_{\mu e}$ of dark photon interaction with muon and electron.}
		\label{Bounds_LFV1G2}
	\end{figure}

For the case of finite non-diagonal couplings, $g^{A,V}_{ik} \neq 0$, and vanished $g^{V,A}_{ii}\equiv 0$, the typical contribution of a massive neutral dark vector 
boson  to $(g-2)$ was calculated explicitly in Ref.~\cite{Kachanovich:2021eqa}. 
In Fig.~\ref{DiagG2} we show the corresponding one-loop diagram. 
These quantities are given by one-dimensional integrals over 
Feynman parameter: 
	\eq
	\delta a^{V}_{l} &=&  \frac{(g_{lf}^V)^{2}}{4\pi^2} \, 
	y_l \, 
	\int\limits_0^1 \, dx \, \frac{1-x}{\Delta(x,y_A,y_l)} \,\Big[ 
	x \ \Big(2 - y_l (1+x)\Big)  \nn
	\\&+&\, \frac{(1-y_l)^2}{2 y_A^2} \, (1 + y_l x) \, (1-x) \Big]\, 
	\,, \label{aV_Rxi}\\ 
	\delta a^{A}_{l} &=&  - \frac{(g_{lf}^{A})^{2}}{4\pi^2} \, 
	y_l \, 
	\int\limits_0^1 \, dx \, \frac{1-x}{\Delta(x,y_A,y_l)} \,\Big[ 
	x \ \Big(2 + y_l (1+x)\Big)  \nn
	\\
	&+&\,\frac{(1+y_l)^2}{2 y_A^2} \, (1 - y_l x) \, (1-x) \Big] 
	\,, \label{aA_Rxi} 
	\en 
where we use the following notations: $y_l = m_l/m_f$, 
$y_A = m_{A'}/m_f$ and $\Delta(x,a,b) = a^2 x + (1-x) (1-b^2 x)$, $m_l$ is mass of external  lepton, $m_f$ is mass of internal 
lepton in loop. In Fig.~\ref{Bounds_LFV1G2} we show the typical bounds on the non-diagonal coupling $g^V_{\mu e}$ from $(g-2)_e$ using 
a set of the ratios $g^A_{\mu e}/g^V_{\mu e}$. Similar bounds we can obtain from $(g-2)_\mu$ discrepancy for $g^V_{\mu \tau}$ (see details in Ref.\cite{Kachanovich:2021eqa}).

The most stringent constraint on $g_{\mu e}^V$ implies
$g_{\mu e}^A \ll 10^{-8}$, since the typical contributions of
vector and axial-vector mediators have an opposite signs in Eqs.~(\ref{aV_Rxi}) and (\ref{aA_Rxi}) and thus
contribution of vector field is maximal. For the
benchmark  ratio $g_{\mu e}^V\simeq g_{\mu e}^A$ the vector
and axial-vector  terms almost compensate each other at $m_{A'} \lesssim 2 m_\mu$,
which leads to a weakening  of the limit on $g_{\mu e}^V$.
Remarkably,  there is a typical sensitivity mass threshold  
at $m_{A'} \gtrsim 2 m_\mu$ for the benchmark ratio $g^A_{\mu e}/g^V_{\mu e}\simeq 1$.
	
\section{Invisible lepton decay mode}
 \label{LFV-two-body-decay}

If dark photon with  non-diagonal LFV couplings to
muons (tau) and electrons (muon) is relatively light, i.~e.~$m_{i} \gtrsim m_{A'}+m_f$, then the decays  $\mu\to e +\mbox{inv.}$, $\tau\to e +\mbox{inv.}$, and 
$\tau\to \mu +\mbox{inv.}$ are kinematically allowed, where we imply the invisible decay of dark photon in the final state $A' \to \chi \overline{\chi}$.  
In addition, the typical invisible charged LFV decays are searched in the processes with lepton flavor conversion accompanied by production of 
pair of neutrino and anti-neutrino~\cite{TWIST:2014ymv,ARGUS:1995bjh}. 
It is worth noticing that the fixed target experiments with missing energy/momenta technique can be a suitable tool to search for them.

The strongest bounds on non-diagonal couplings of dark photon 
with leptons for area of $m_{A'} < 2m_e$ are obtained from experimental searches 
for the decays $\tau \to \mu + \mbox{inv.}$ and $\tau \to e + \mbox{inv.}$ 
by the ARGUS Collaboration~\cite{ARGUS:1995bjh}, and recent data from 
the Belle-II Collaboration~\cite{Belle-II:2022heu}. 
We use the PDG data~\cite{Workman:2022ynf} to constrain the non-diagonal 
dark photon-lepton couplings $g^{V,A}_{\tau e}$ and $g^{V,A}_{\tau \mu}$. 
For axion-like particles (ALPs) the bounds were obtained in Ref.~\cite{Bauer:2019gfk}. For $\mu \to e +\mbox{inv.}$ decay we can use the limit obtained by 
the TWIST Collaboration~\cite{TWIST:2014ymv} predicts branching 
ratio at level up to $5.8 \times 10^{-5}$. Using this constraint, 
we show exclusion plots in Figs.~\ref{Bounds_LFV_NA64and_etc_V} 
and~\ref{Bounds_LFV_NA64and_etc_A}.

Decay widths of such rare charged LFV decays $l_i \to l_f + A'$ 
in vector and axial-vector cases are defined as 
\eq
& &\Gamma(l_i \to l_f + A')_{V/A} = 
\frac{3 \Big(g_{if}^{V/A}\Big)^2 m_i}{8 \pi} 
\, \lambda^{1/2}(y_i^2,y_f^2,1)
\nonumber\\
&\times&\Big[(y_i \mp y_f)^2 - 1 + \frac{y_i^2}{3} 
\, \lambda(y_i^2,y_f^2,1) \Big] \,,
\en
where $y_i=m_i/m_{A'}$, $y_f=m_f/m_{A'}$, and
$\lambda(x,y,z) = x^2 + y^2 + z^2 - 2 x y - 2 x z - 2 y z$ 
is the K\"allen kinematical triangle function. 

It is important to note that this decay depends only on
non-diagonal coupling, while other LFV lepton decays 
as $l_i\to l_f +\gamma$ or $l_i \to 3l_f$ have a dependence 
from the product of diagonal and non-diagonal couplings. Analysis of the parameter space of dark photon was done in Ref.~\cite{Heeck:2016xkh} where implication of 
the charged LFV two body decay has been noted.

\section{The fixed-target experiments}
 \label{FixTarget}

Fixed target experiments represent themselves very useful and 
crucial test of physics of feebly interacting particles and make 
up a third of the experimental base for searching and 
analysis of DM~\cite{Lanfranchi:2020crw} or hidden sector.    
In this section we study dark photon emission reactions 
with change of lepton flavor. 

In the framework of the benchmark scenario (\ref{eq:A'-psi-1}), stringent limits 
can be established from the missing energy/momenta experiments by analysis of $l_i+N\to l_f+N+A'$ process. 
In particular, we will discuss the potential for the 
fixed target experiments with lepton beams such as 
NA64$e$~\cite{Gninenko:2017yus,Gninenko:2019qiv,%
Banerjee:2019pds,%
NA64:2021xzo,Andreev:2021fzd,Arefyeva:2022eba},
LDMX~\cite{Berlin:2018bsc,LDMX:2018cma,Ankowski:2019mfd,%
Schuster:2021mlr,Akesson:2022vza}, 
NA64$\mu$~\cite{Gninenko:2014pea,Gninenko:2018tlp,%
Kirpichnikov:2021jev,Sieber:2021fue} and
$\mbox{M}^3$~\cite{Kahn:2018cqs,Capdevilla:2021kcf}, 
which can be used to probe the invisible signatures associated with 
a lepton flavor violation and missing energy process. Missing energy signals can be evidence
of dark photon emission that involves LFV processes. 
Analysis of final lepton states can give information about specific 
LFV channels of lepton conversion with dark photon emission.

The existed (NA64$e$~\cite{Gninenko:2017yus,Gninenko:2019qiv,Banerjee:2019pds,
NA64:2021xzo,Andreev:2021fzd,Arefyeva:2022eba} and NA64$_\mu$~\cite{Gninenko:2014pea,Gninenko:2018tlp,
Kirpichnikov:2021jev,Sieber:2021fue}) and future experiments 
(LDMX~\cite{Berlin:2018bsc,LDMX:2018cma,Ankowski:2019mfd,Schuster:2021mlr,Akesson:2022vza}, 
$\mbox{M}^3$ \cite{Kahn:2018cqs,Capdevilla:2021kcf})
are noted by us due to a possibility of using missing energy/momentum 
technique. Experiments NA64$_e$ and LDMX use electron beams which collide to 
active lead and aluminium targets, respectively. Experiments NA64$_\mu$ and 
$\mbox{M}^3$ use muon beams, $\mbox{M}^3$ will use tungsten target. We collect the main parameters of the fixed-target 
experiments NA64$_e$, NA64$_\mu$, M$^3$, and LDMX in Table~\ref{ParamTable}.

It is worth noticing that analysis of LFV with dark matter emission at fixed-target experiments was done for the scalar case.
In this respect we can mention Refs.~\cite{Gninenko:2022ttd,Radics:2023tkn,Ema:2022afm} that examine the invisible scalar portal scenarios.  
A possibility of $e-\tau$ and $\mu-\tau$ conversion in deep-inelastic lepton scattering off nuclei
was proposed and studied in detail in Ref.~\cite{Gninenko:2018num}
based on assumption of local four-fermion lepton-quark interaction. Besides, huge research lepton conversion  on nucleons was made in Refs.\cite{Gonzalez:2013rea,Faessler:2005hx,Faessler:2004ea,Faessler:2004jt}.    

\subsection{Signal of lepton conversion}
\label{miss-energy}

\begin{figure}[b!]
	\includegraphics[width=0.48\textwidth, trim={4.5cm 11.5cm 4.5cm 11.7cm},clip]{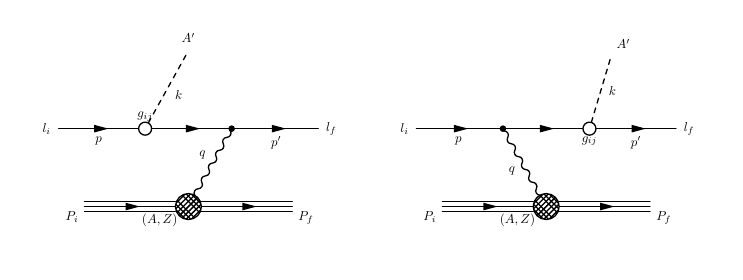}
	\caption{Lowest-order diagrams describing LFV emission of dark photon $A'$ in lepton scattering on fixed atomic target.}
	\label{Bounds_FixTargetDiag}
\end{figure}

Based on the setup of the missing energy experiment we will consider the reaction 
of lepton scattering off the nucleus target as shown in Fig.~\ref{Bounds_FixTargetDiag}. We propose that dark photon have a lifetime larger than time of flight inside the detector or dominant decay mode into dark matter. The invisible two-body 
decay width of dark photon $A'$ into $\bar{\chi}\chi$ pair is given by
\eq 
\Gamma_{A'\to \bar{\chi}\chi } = \frac{g_D^2}{12 \pi} 
\, m_{A'} \, 
(1 + 2 y_\chi^2) \, (1 - 4 y_\chi^2)^{1/2} \,.  
\en 
where $g_D$ is coupling dark photon 
with dark fermions and $y_\chi = m_\chi/m_{A'}$.

For calculation we will use Weizs\"acker-Williams (WW) 
approximation~\cite{Budnev:1975poe,Tsai:1973py}. 
In this case, the signal of
interaction of incoming leptons ($e$, $\mu$) with atomic target 
can be effectively described through the Compton-like conversion on 
virtual photons $\gamma^*$, i.~e. via $l \gamma^* \to l' A'$.  
We also suppose that target nucleus has a spin$=1/2$ and 
its coupling to photon is  $i e Z F(t) \gamma_\mu$, where 
$F(t)$ is an elastic form factor depending on $t=-q^2>0$ (nucleus transfer momentum squared), $Z$ is the charge of nucleus. $F(t)$ has the form 
\eq 
F(t)=\frac{a^2 t}{(1+a^2 t)}\frac{1}{(1+t/d)} \,,
\en

\onecolumngrid
\onecolumngrid
\begin{center}
\begin{table}[t]
	\centering
	\caption{Parameters of the fixed-target experiments NA64$_e$, NA64$_\mu$, M$^3$,  and LDMX: parameters of target ($A, Z$), first radiation length $X_0$, effective thickness of the target ($L_T$), energy of scattering beam and current (started) and planned accumulate of leptons on target, $x_{\text min}$ characterizing 
    a window of search. 
 }
	\begin{tabular}[t]{rccccccccc}
 \hline
		\hline
		 & $e$-conv.	& $\mu$-conv. &$A$\, ($Z$) & $E$ (GeV)& $\rho$ ( g cm$^{-3}$)& $X_0$ (cm)& $L_T$ (cm)& $x_{\text min}$ & LoT (projected LoT)\\
		\hline
	   NA64$_e$:& $e N\to   N A' \mu (\tau) $& -- &207 (82)&100 & 11.34 & 0.56 &0.56 &0.5 & 3.3$\times  10^{11}$ \quad($5 \times 10^{12}$)\\
		   	NA64$_\mu$:& -- & $\mu N\to   N A' e (\tau)$  &207 (82)&160 & 11.34 & 0.56& 22.5 & 0.5 & $10^{10}$ \quad ($10^{13}$ )\\
		M$^3$:&  -- & $\mu N\to   N A' e (\tau)$  &184 (74)&15&19.3&0.35& 17.5&0.4& $10^{10}$ \quad ($10^{13}$)\\
		LDMX: & $e N\to   N A' \mu (\tau) $& -- &27 (13)&16&2.7&8.9&3.56 &0.7&$10^{16}$ \quad ($10^{18}$) \\
		\hline
  \hline
  \label{ParamTable}
	\end{tabular}
\end{table}%
\end{center}
\twocolumngrid
\twocolumngrid
where $a=111 Z^{-1/3}/m_e$ and 
$d=0.164 A^{-2/3}\, \mbox{GeV}^2$ are the 
screening and nucleus size parameters, 
respectively~\cite{Bjorken:2009mm}. These parameters of nuclear form factor include $m_e$ which is mass of electron and $A$ is atomic weight number.

Differential cross section for $2\to 3$ process, presented in Fig.~\ref{Bounds_FixTargetDiag},
in the framework of the WW approximation is given by 
\begin{equation}
\frac{d \sigma_{l Z \to l' Z A'}}{d(pk ) 
d (k \mathcal{P}_i) } \simeq      
\frac{\alpha \gamma_Z }{\pi (p' \mathcal{P}_i)} 
\cdot \frac{d \sigma_{l  \gamma^* \to l' A'}}{d(pk)} 
\Bigl|_{t=t_{min}} \,,
\label{dsdxdthetaLeptophilic1}
\end{equation}
where $t_{min}$ is a minimal momentum transfer that is provided below in 
Eq.~(\ref{tminDefinition1}), $\alpha\simeq 1/137.036$ is a fine structure constant, 
$\gamma_Z$ is the effective photon flux from nucleus defined as
\eq
\gamma_Z=Z^2\int\limits^{t_{max}}_{t_{min}} 	
dt \frac{t-t_{min}}{t^2}F^2(t).  
\label{flux}
\en 
where $\mathcal{P}_i$ is momentum of nuclear, $p$, $p'$ and $k$ are momenta of initial, finale leptons and dark photon (see definition in Fig.~\ref{Bounds_FixTargetDiag}), $t_{max}$ and $t_{min}$ are kinematic bounds is obtained from the energy-conserving $\delta$-function in the
Lorentz-invariant phase space (details in Ref.\cite{Liu:2017htz}).

By using such approximation, we can calculate the typical number of 
missing energy events for experiments with  
lepton beam impinging on the fixed target   
\begin{equation}
N_{A'} \simeq \mbox{LoT}\cdot \frac{\rho N_A}{A} L_T \int\limits^{x_{max}}_{x_{min}}
dx \frac{d \sigma_{2\to3}}{dx}(E)\,,
\end{equation}
where $N_A$ is the Avogadro's number, $\mbox{LoT}$ is number of leptons accumulated on target, $\rho$ 
is the target density, $L_T$ is the effective interaction length 
of the lepton in the target, $d\sigma_{2\to3}/dx$ is the 
differential cross section of the lepton conversion  $l N \to l' N A'$, $E$ is a initial lepton beam energy, $x=E_{A'}/E$ is the energy fraction that dark 
photon carries away, $x_{min}=E_{\text cut}/E$ and $x_{max}\simeq 1$
are the minimal and maximal fraction of dark photon energy respectively
for the  regarding  experimental setup, $E_{\text cut}$ is a detector missing energy cut that is determined by the specific facility. The typical parameters of the experiments can be found in 
Refs.~\cite{Zhevlakov:2022vio,Kirpichnikov:2021jev,%
Gninenko:2017yus,Gninenko:2019qiv,Banerjee:2019pds,%
NA64:2021xzo,Andreev:2021fzd,Berlin:2018bsc,LDMX:2018cma,%
Ankowski:2019mfd,Schuster:2021mlr,Akesson:2022vza,
Sieber:2021fue,Kahn:2018cqs,Capdevilla:2021kcf}. For muon beam experiments (NA64$_\mu$ and M$^3$) we will used definition $\mbox{MoT}$ for number of muons on target, for electron beam experiments (NA64$_e$ and LDMX) will be used  $\mbox{EoT}$ for number of electrons on target. There are the parameters 
determined by technical characteristics of detectors and by background cutoff. 

\subsection{Cross-section}
\label{cross_section}

Amplitude of the $2\to 2$ process $l_i(p)+\gamma(q) \to l_f(p')+A'(k) $ is given by 
\begin{align}
M^{2\to2}\!\!\!\!\!=&i e\, \epsilon^\mu_\lambda  \epsilon^{*\alpha}_{\lambda'}   \bar{u}_f(p',s')\Bigg[ \gamma_\mu \frac{\not\!{p}-\not\!{k}+m_i}{\tilde{u}}(\gamma_\alpha  g^V_{if}+\gamma_5\gamma_\alpha g^A_{if}) \nn \\
&+(\gamma_\alpha  g^V_{if}+\gamma_5\gamma_\alpha g^A_{if}) \frac{\not\!{p'}+\not\!{k}+m_f}{\tilde{s}}\gamma_\mu
\Bigg] u_i(p,s),
\end{align}
where used notation $\not\!{p}=\gamma^\mu p_\mu$, $g^V_{if}$ and $g^A_{if}$ are the non-diagonal couplings of interaction of dark photon with leptons, $\epsilon$ is the polarization vector of photon or dark photon, $m_i$ and $m_f$ are the masses of initial and final leptons. The sums over polarizations are given by 
\eq
\sum_\lambda  \epsilon^\mu_\lambda \epsilon^{*\nu}_\lambda =  - g^{\mu\nu},
\en
for visible photon and
\eq
\sum_{\lambda'}  \epsilon^\mu_{\lambda'} \epsilon^{*\nu}_{\lambda'} = - g^{\mu\nu}+		\frac{k^{\mu}k^{\nu}}{m_{A'}^2},
\en
for massive dark photon. In case of visible photon 
we use $\xi\to \infty$ gauge omitting ghost fields contribution. 

\newpage

\onecolumngrid
\onecolumngrid
\begin{figure}[t!]
\includegraphics[width=0.99\textwidth,trim={4.5cm 12.6cm 4.6cm 12.8cm},clip]{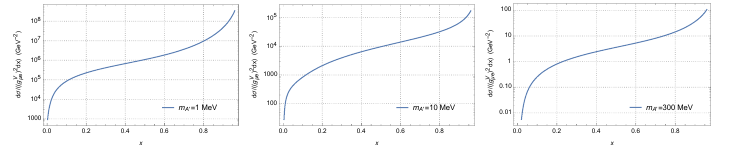}
\caption{Differential cross section $d\sigma/((g^V_{\mu e})^2dx)$ for the process of electron to muon conversion in the case of vector non-diagonal coupling of dark photon with lepton for the NA64e experiment with different masses of dark photon and the typical energy of electron beam $E=100$ GeV. }
\label{E_mu_dsigma_dx1}
\end{figure}
\begin{figure}[t!]
	\includegraphics[width=0.99\textwidth,trim={4.5cm 12.6cm 4.6cm 12.8cm},clip]{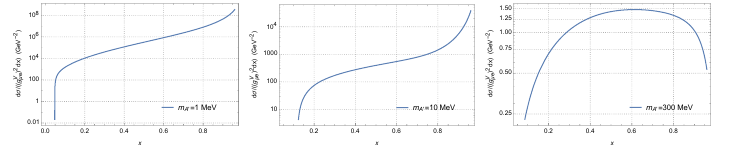}
\caption{Differential cross section $d\sigma/((g^V_{\mu e})^2dx)$ for the process of muon to electron conversion in the case of vector non-diagonal coupling of dark photon with lepton 
for the NA64$\mu$ experiment with different masses of dark photon and the typical energy of muon beam $E=160$ GeV.}
\label{E_mu_dsigma_dx2}
\end{figure}
\twocolumngrid

After averaging and summation over spins of leptons  
and polarizations of vector bosons one gets the matrix element squared 
\eq
\overline{|M^{2\to2}|^2} &=& \frac{1}{4} \sum_{s,s'} \, 
\sum_{\lambda,\lambda'}|M^{2\to2}|^2 \nn\\ 
&=&e^2 (g^V_{if})^2 A^{2\to2}_V+e^2 (g^A_{if})^2 A^{2\to2}_A.
\en 
In our calculation we will use the so-called modified Mandelstam 
variables: 
\eq
\tilde{s} &=& (p'+k)^2-m_f^2=2(p'k) +m_{A'}^2\,, \nn\\
\tilde{u} &=& (p-k)^2-m_i^2=-2(pk) +m_{A'}^2\,,  \\
t_2 &=& (p'-p)^2=-2(p'p) +m_{i}^2 +m_{f}^2\,,    \nn\\
t&=&q^2, \nn
\en	
which satisfy the condition $\tilde{s}+t_2+\tilde{u} = m_{A'}^2$.  

When considering the process, we have an energy of initial lepton energy much greater than masses of $m_{A'}$ and $m_f$. In this case we can use WW approximation and can propose with high reliability that the final state of lepton and dark photon are highly collinear.
When, in the small-angle approximation (which means that $t=t_{\text min}$) we imply that $\bf{q}$ and $\bf{V}=\bf{k}-\bf{p}$ are collinear~\cite{Tsai:1973py,Liu:2017htz} and
the modified Mandelstam variables are given by
\eq	
U=-\tilde{u}&\approx&E^2x \theta^2+m_i^2x+\frac{1-x}{x}m_{A'}^2,	\nn\\
\tilde{s}&\approx&\frac{U+(m_f^2-m_i^2)x}{1-x},\\
t_2&\approx&-\frac{x}{1-x}\left(U+(m_f^2-m_i^2)\right)+m_{A'}^2,\nn\\
t_{min}&\approx&\frac{(\tilde{s}+(m_f^2-m_i^2))^2}{4E^2} \label{tminDefinition1}
\,.
\en	
Note that if we set $m_i=m_f$, then  we reproduce formulas 
presented in Ref.~\cite{Bjorken:2009mm,Liu:2017htz}. 

The double differential cross-section can be presented in the form
\begin{equation}
\frac{d \sigma_{2\to 3}}{dx \, d\cos\theta_{A'}} 
\simeq \frac{\alpha \gamma_Z }{\pi(1-x)}\cdot  E^2 x \beta_{A'} 
I_{\tilde{s}} \cdot \frac{d \sigma_{2\to 2}}{d (pk)}\,,
\label{dsdxdcostheta2}
\end{equation}
where $\beta_{A'}=(1-m_{A'}^2/(x E)^2)^{1/2}$ is the 
velocity of dark photon in the laboratory frame, 
$I_{\tilde{s}}=\tilde{s}^2/(\tilde{s}+(m_f^2-m_i^2))^2 \beta_{m_i}^{-1}$, 
$\beta_{m_i}=\sqrt{1-m_i^2/E^2}$. Full analytical expression
for the effective photon flux $\chi$ is presented in the Ref.~\cite{Kirpichnikov:2021jev}. It is known that 
the elastic form-factor $G_{\text el}(t)$ is proportional to 
$\propto Z^2$. An inelastic form-factor is suppressed by a factor $Z$, 
i.e. $G_{\text inel}(t)\propto Z$, and therefore 
for the heavy target nuclei $Z\propto \mathcal{O}(100)$ one 
can safely ignore it in calculation. 

\newpage
\onecolumngrid
\onecolumngrid
\begin{figure}[t!]
\includegraphics[width=0.99\textwidth,trim={4.5cm 12.4cm 4.6cm 12.5cm},clip]{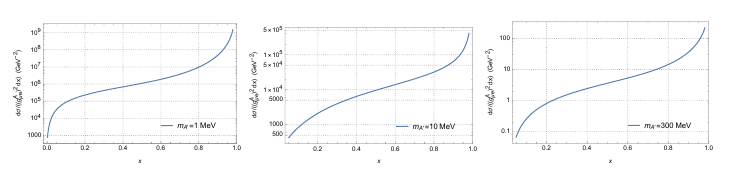}
\caption{The same as in Fig.~\ref{E_mu_dsigma_dx1} but for axial vector field.}
	\label{E_mu_dsigma_dx3}
\end{figure}

\begin{figure}[t!]
\includegraphics[width=0.99\textwidth,trim={4.5cm 12.4cm 4.6cm 12.5cm},clip]{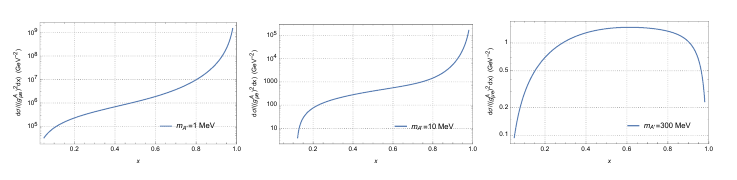}
	\caption{The same as in Fig.~\ref{E_mu_dsigma_dx2} but for axial vector field.}
\label{E_mu_dsigma_dx4}
\end{figure}
\twocolumngrid

Differential cross section~\cite{Liu:2017htz} can be rewritten as
\eq
\frac{d\sigma^{V,A}_{2\to2}}{d(pk)} = 2\frac{d\sigma}{dt_2} = \frac{\overline{|M^{2\to2}|}^2}{8\pi \tilde{s}^2} = (g^{V,A}_{if})^2 \frac{\alpha}{\tilde{s}^2} A_{V,A}^{2\to2}
\,,
\label{dsigmadpk}
\en
where the vector and axial-vector parts of the amplitudes squared can be written, respectively, as:
\onecolumngrid
\eq
A^{2\to2}_{V,t=t_{\text min}}&\approx& \frac{1}{\tilde{s}^2 U^2 m_{A'}^2}\Big(\tilde{s} (\tilde{s} - U)^2 U (m_i- m_f)^2 
+ (2 \tilde{s} U (\tilde{s}^2 + U^2) + (\tilde{s}^2 + U^2) 
m_i^4+ 6 (\tilde{s} - U)^2 
m_i^3 m_f +  \nn \\
&+&2 (\tilde{s}^3 + 2 \tilde{s}^2 U + U^3) 
m_f^2+ (\tilde{s}^2 + U^2) 
m_f^4 + 
6 m_i m_f (U-\tilde{s} ) (2 \tilde{s} U + (U-\tilde{s} ) 
m_f^2)  \nn \\ &-& 2 
m_i^2(\tilde{s}^3 + 2 \tilde{s} U^2 + 
U^3 + (3 \tilde{s}^2 - 4 \tilde{s} U + 3 U^2) 
m_f^2)) 
m_{A'}^2 + 
2 (2 \tilde{s} U (U-\tilde{s}) + U (-3 \tilde{s} + 2 U) 
m_i^2 + \nn \\
&+&6 \tilde{s} U m_i m_f + \tilde{s} (2 \tilde{s} - 3 U) 
m_f^2) 
m_{A'}^4 + 4 \tilde{s} U 
m_{A'}^6\Big),
\en
\eq
A^{2\to2}_{A,t=t_{\text min}}&\approx& \frac{1}{\tilde{s}^2 U^2 m_{A'}^2}	\Bigg(\tilde{s} (\tilde{s} - U)^2 U (m_i + m_f)^2 
+ (2 \tilde{s} U (\tilde{s}^2 + U^2) + (\tilde{s}^2 + U^2) 
m_i^4- 6 (\tilde{s} - U)^2 
m_i^3 m_f  \nn\\ &+& 
2 (\tilde{s}^3 + 2 \tilde{s}^2 U + U^3) 
m_f^2 + (\tilde{s}^2 + U^2) 
m_f^4 + 
6 ( U-\tilde{s}) m_i m_f(-2 \tilde{s} U + (\tilde{s} - U) 
m_f^2) \nn \\ &-& 2 
m_i^2(\tilde{s}^3 + 2 \tilde{s} U^2 + 
U^3 + (3 \tilde{s}^2 - 4 \tilde{s} U + 3 U^2) 
m_f^2)) 
m_{A'}^2 - 
2 (2 \tilde{s} (\tilde{s} - U) U + (3 \tilde{s} - 2 U) U 
m_i^2 \nn \\ &+& 
6 \tilde{s} U m_i m_f + \tilde{s} (  3 U-2 \tilde{s}) 
m_f^2) 
m_{A'}^4 + 4 \tilde{s} U 
m_{A'}^6\Bigg).
\en 
\twocolumngrid

In our calculations we used the 
FeynCalc package~\cite{Shtabovenko:2016sxi,Shtabovenko:2016whf,Shtabovenko:2020gxv} 
and the following assumptions to exploit the WW approximation: 
(i) cross section for processes with $m_i > m_f$
was evaluated by taking into account  
non-zero emission angle of the final particle~\cite{Byckling:1971vca}: 
$\theta_{\text min}\simeq  [(m_i^2+m_{A'}^2 -m_f^2)/(|\mathbf{k}||\mathbf{p}|)]^{1/2}$, (ii) the maximal typical angle of dark photon emission was set to be 
$\theta_{\text max} \simeq  0.1$.
As a result, the integration over the angle $\theta_{A'}$ is performed in the range $\theta_{\text min} \lesssim \theta_{A'} \lesssim \theta_{\text max}$. 

In order to illustrate  our results for the  differential cross sections $d\sigma/dx$ 
we make a comparison for different combinations of initial and final states 
of leptons. In particular, we demonstrate it 
in Figs.~\ref{E_mu_dsigma_dx1}-\ref{E_mu_dsigma_dx4}. 
For the case of equal masses of initial and final leptons we 
reproduce the numerical results presented in Refs.~\cite{Liu:2017htz,Kirpichnikov:2020tcf}. 
These cross sections $d\sigma/dx$ are 
calculated by using Eqs.~(\ref{dsdxdcostheta2}) and~(\ref{dsigmadpk}) 
implying the  lepton beam energy and target characteristics of the NA64e 
and NA64$_\mu$ experiments. One can see the similar  shapes of the vector and axial-vector dark photon in case of $e\to \mu$ conversion cross sections at $m_{A'} \simeq 1 \, \mbox{MeV}$. 
For larger masses $m_{A'} \gtrsim 300\, \mbox{MeV}$ the differential 
spectra of vector and axial vector are almost coincided. 
For the case of  $\mu \to e$ conversion cross sections one can see 
there are sharp peaks at  $x\simeq 1$ for the relatively light dark photon $m_{A'} \lesssim 10\, \mbox{MeV}$. The corresponding peaks are mitigated for heavy masses $m_{A'} \gtrsim 300\, \mbox{MeV}$.

\vspace{\columnsep}
\begin{figure}[b!]
	\centering
	\includegraphics[width=0.48\textwidth, trim={5cm 5.3cm 8cm 5cm},clip]{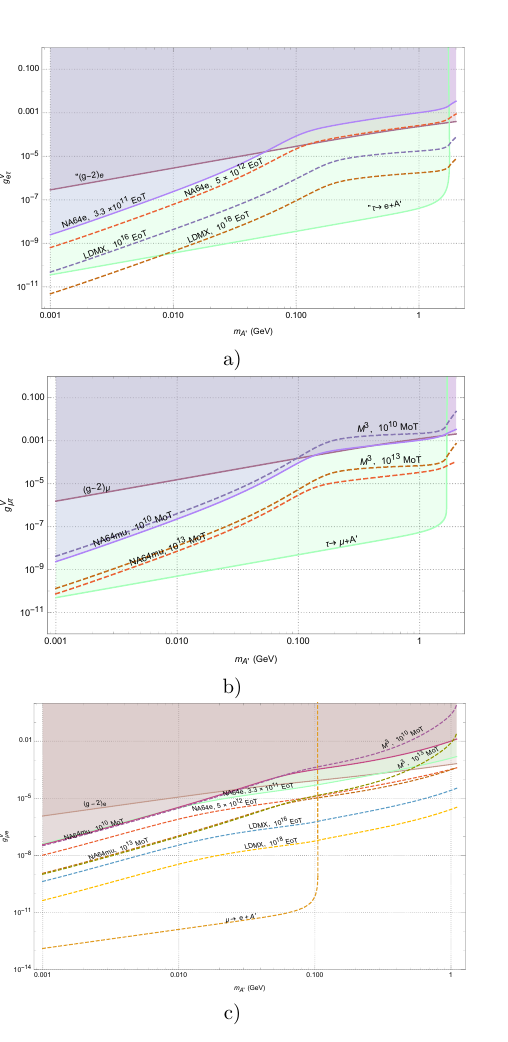}
	\caption{Bounds on vector non-diagonal coupling  
		of dark photon with leptons using characteristics of running and proposed 
		fixed-target experiments NA64e, NA64$_\mu$, LDMX, and M$^3$. Bounds 
		from $g-2$ of leptons and invisible lepton LFV decay 
		$l_i \to l_f +A'$ are included.  Panel (a): bounds for $g^V_{e \tau }$,  
		panel (b): bounds for $g^V_{\mu \tau }$, panel (c): bounds for $g^V_{\mu e}$.}
	\label{Bounds_LFV_NA64and_etc_V}
\end{figure}
\twocolumngrid

\section{Results and discussion}
\label{summary}

In this section we derived the bounds on non-diagonal $g^{V,A}_{ik}$ couplings of dark photon with leptons from the lepton scattering experiments at fixed targets, that implies LFV  conversion $l N \to l' N A'$ followed 
by the  invisible  decay of dark photon with  $\mbox{Br} (A'\to \chi \overline{\chi}) \simeq 1$.  We set 
the  diagonal couplings to be $g^{V,A}_{ii}\equiv 0$ throughout the analysis.   

\begin{figure}[t]
    \centering
    \includegraphics[width=0.48\textwidth,  trim={5cm 5.3cm 8cm 5cm},clip]{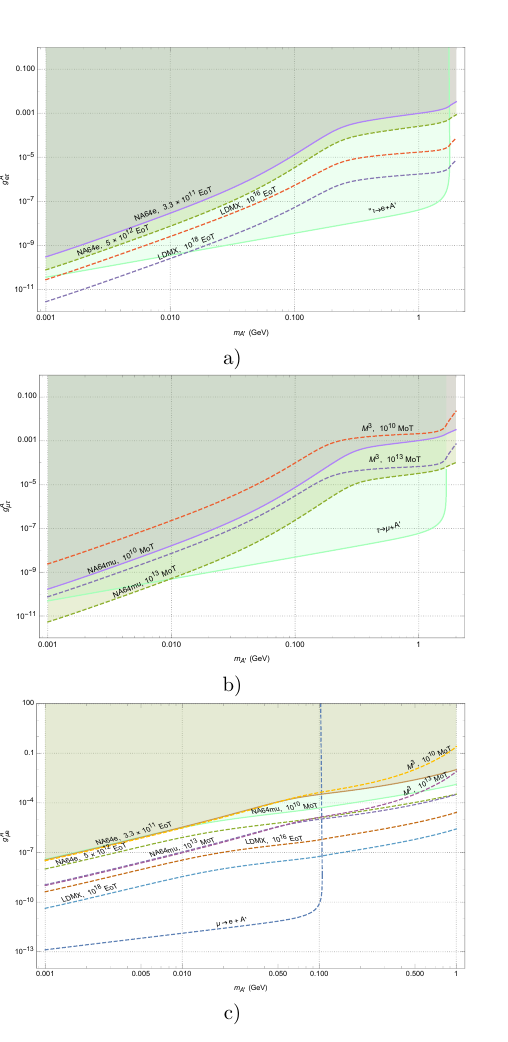}
 \caption{Bounds axial-vector for non-diagonal coupling interaction of dark photon with leptons using current and proposal experimental characteristics of fix target  experiments NA64e, NA64$_\mu$, LDMX and M$^3$. Bounds from $g-2$ of leptons and invisible lepton LFV decay $l_i \to l_f +A'$ are included.  a) Bounds for $g^A_{e\tau }$,  b) bounds for $g^A_{\mu \tau }$ and  c) bounds for $g^A_{\mu e}$. }
\label{Bounds_LFV_NA64and_etc_A}
\end{figure}

Our results for exclusion limits on vector and axial-vector couplings
are presented in Fig.~\ref{Bounds_LFV_NA64and_etc_V} and Fig.~\ref{Bounds_LFV_NA64and_etc_A} at $90~\%\,\mbox{C.L.}$,
implying zero observed signal events and background free case  and
null-result of the fixed target experiments and on assumption that theoretical uncertainty is 
negligible, $N_{A'}\lesssim 2.3$. In particular, the analysis reveals~\cite{NA64:2023wbi} that  the 
electron beam  mode  NA64$_e$  is background free for 
$\mbox{EoT}\simeq \mathcal{O}(1)\times 10^{12}$.   In addition, for our analysis of the muon beam mode at
NA64$_\mu$ we rely on the study~\cite{Radics:2023tkn,Sieber:2023nkq,Sieber:2021fue} that implies the 
background  suppression at the level of  $\lesssim \mathcal{O}(1) \times 10^{-13}$ per muon
for the LFV process due to the  emission of spin-0 boson $\phi$, i.~e.~in the reaction 
$l N \to l' N \phi$. Furthermore, for the background suppression of both LDMX and $\mbox{M}^3$ facilities 
we  rely on the explicit analysis of~\cite{LDMX:2019gvz} and~\cite{Kahn:2018cqs} 
respectively for the reactions without LFV, $l N \to l N A'$, implying conservatively that the regarding  background  rejections would be the same for LFV process, $l N \to l' N A'$. 

To search for a channel of LFV conversion in $l N \to l' N A'$ process is needed to search a signal with missing energy and single different flavor lepton.  For the benchmark conversion
$e (\mu) N \to  \tau  N A'$ and the typical
dark photon mass range $m_{A'} \lesssim 1\, \mbox{GeV}$,
the existing fixed-target experiments (NA64$_e$ and NA64$_\mu$) can not reach the  bound on $\tau \to e (\mu) A'$ from ARGUS experiment~\cite{ARGUS:1995bjh} for both vector
 $g^V_{e \tau  } (g^V_{\mu \tau  })$  and  axial-vector $g^A_{e \tau  } (g^A_{\mu \tau  })$  couplings. For the considered benchmark scenarios, the most stringent constraints on the couplings
$g^{V}_{e \tau } \lesssim 10^{-11}$ and $g^{A}_{e\tau } \lesssim 10^{-11}$ at $m_{A'} \lesssim 1\, \mbox{MeV}$ are expected from the projected  statistics  ($\mbox{EoT}=10^{18}$) of the LDMX experiment.
However, for the typical mass range $m_{A'} \gtrsim 100\, \mbox{MeV}$ the LDMX is able to set the constraint at the level of $g^{V,A}_{\mu e} \lesssim 10^{-7}$. Remarkably,
for the typical mass range $m_{A'} \gtrsim m_\tau \simeq 1.7 \, \mbox{GeV}$, the LDMX facility can set the bounds at
$g^{V,A}_{e \tau } \lesssim 10^{-5} - 10^{-4}$. For the accumulated statistics  of NA64$_e$ at the level of $\mbox{EoT}\simeq 3.3 \times 10^{11}$ the corresponding experiment provides the bound $g_{e\tau}^A \lesssim 5\times 10^{-3}$ for the mass threshold range
$m_{A'} \gtrsim 1.7~\mbox{GeV}$. In addition, the NA64$_e$ also rules out the typical regions
$g_{\mu e}^V \lesssim 10^{-4}$ for $100~\mbox{MeV} \lesssim m_{A'} \lesssim 300~\mbox{MeV}$
and $g_{\mu e}^A \lesssim 10^{-3}-10^{-2}$ for
$100~\mbox{MeV}~\lesssim m_{A'}~\lesssim~1~\mbox{GeV}$. 

The NA64$_\mu$ fixed target experiment can provide a more stringent limit than the M$^3$ facility.
Moreover, the expected reaches of the LDMX and  NA64$_\mu$ experiments at the masses of
dark photon $m_{A'} \lesssim 1\, \mbox{MeV}$  can be comparable with the current bounds from two-body LFV lepton decay, $\tau \to \mu (e) A'$.
In addition, the current limits from missing energy data of the NA64$_e$ and NA64$_\mu$
experiments are comparable with  the bounds from $(g-2)_{e,\mu}$ tensions of  leptons.  
We note that for the projected statistics $\mbox{MoT}\simeq 10^{13}$ both NA64$_\mu$ and M$^3$
can rule out  the typical region $g_{\mu \tau}^{V,A} \lesssim 10^{-4} - 10^{-3}$
for  the relatively heavy masses of dark photon $m_{A'} \gtrsim 1.7 \, \mbox{GeV}$. Moreover,
for the mass range $100 \, \mbox{MeV} \lesssim  m_{A'} \lesssim 1\, \mbox{GeV}$ the NA64$_\mu$
is able to set the constraint $ g^{A,V}_{ \mu e } \lesssim 10^{-5}$ for $\mbox{MoT}\simeq 10^{13}$.  

Finally, we would like to note that one can probe the resonant production of dark 
photons in the LFV reaction of muon scattering off atomic electrons 
$\mu^+ e^- \to A' \to \chi \bar{\chi}$ at both NA64$_\mu$ and  M$^3$ muon beam fixed target experiments. However, this analysis is beyond the scope of the present paper and we plan to study it in future.

\begin{acknowledgments} 
	The work of A.~S.~Zh. on exclusion limits calculation for the fixed target experiments is supported by Russian Science Foundation
	(grant No. RSF 23-22-00041).  The work of A.~S.~Zh. on exclusion limit calculation from $(g-2)$ tension  is  supported by the Foundation for the Advancement of Theoretical Physics and Mathematics
	"BASIS". 
	The work of D.~V.~K. on signal yield calculation for the $e\to \mu$  conversion at NA64$_e$ experiments  is  supported
	by the  Russian Science Foundation (grant No. RSF 21-12-00379). 
	The work of V.~E.~L. was funded by ANID PIA/APOYO 
	AFB220004 (Chile), by FONDECYT (Chile) under Grant No. 1230160, 
	and by ANID$-$Millen\-nium Program$-$ICN2019\_044 (Chile).

\end{acknowledgments}

\end{document}